\documentclass[aip,superscriptaddress,preprint]{revtex4-1}
\usepackage{graphicx,amsfonts,amssymb,amsmath,hyperref}
\usepackage{dcolumn}
\usepackage{bm}

\newcommand{\Sec}[1]{Sec.\;\ref{#1}}

\newcommand{\be}{\begin{equation}}
\newcommand{\ee}{\end{equation}}
\newcommand{\bea}{\begin{eqnarray}}
\newcommand{\eea}{\end{eqnarray}}
\newcommand{\bsube}{\begin{subequations}}
\newcommand{\esube}{\end{subequations}}
\newcommand{\Eq}[1]{Eq.\,(\ref{#1})}
\newcommand{\Eqs}[1]{Eqs.\,(\ref{#1})}
\newcommand{\Fig}[1]{Fig.\,\ref{#1}}

\begin{document}

\title{Quantum Efficiency of Intermediate-Band Solar Cells Based on Non-Compensated n-p Codoped TiO$_2$}

\author{Fengcheng Wu}
\affiliation{
ICQD, Hefei National Laboratory for Physical Sciences at Microscale,University of Science and Technology of China, Hefei, Anhui, 230026, China
}%
\affiliation{Department of Physics, The University of Texas at Austin, Austin, Texas 78712, USA}
 \author{Haiping Lan}%
\affiliation{
ICQD, Hefei National Laboratory for Physical Sciences at Microscale,University of Science and Technology of China, Hefei, Anhui, 230026, China
}%
\author{Zhenyu Zhang}
\affiliation{
ICQD, Hefei National Laboratory for Physical Sciences at Microscale,University of Science and Technology of China, Hefei, Anhui, 230026, China
}%
\affiliation{School of Engineering and Applied Science, Harvard University, Cambridge, MA 02138, USA}%
\affiliation{Department of Physics, The University of Texas at Austin, Austin, Texas 78712, USA}

\author{Ping Cui}
 \email{cuipg@ustc.edu.cn.}
\affiliation{
ICQD, Hefei National Laboratory for Physical Sciences at Microscale,University of Science and Technology of China, Hefei, Anhui, 230026, China
}%

\date{\today}

\begin{abstract}
As an appealing concept for developing next-generation solar cells, intermediate-band
solar cells (IBSCs) promise to drastically increase the quantum efficiency of photovoltaic
conversion. Yet to date, a standing challenge lies in the lack of materials suitable
for developing IBSCs. Recently, a new doping approach, termed non-compensated n-p codoping,
has been proposed to construct intermediate bands (IBs) in the intrinsic energy band gaps
of oxide semiconductors such as TiO$_2$. We explore theoretically the optimal quantum
efficiency of IBSCs based on non-compensated n-p codoped TiO$_2$ under two different design
schemes. The first preserves the ideal condition that no electrical current be
extracted from the IB. The corresponding maximum quantum efficiency
for the codoped TiO$_2$ can reach 52.7\%. In the second scheme, current is also extracted
from the IB, resulting in a further enhancement in the maximum efficiency to 56.7\%.
Our findings also relax the stringent requirement that the IB location be close to the
optimum value, making it more feasible to realize IBSCs with high quantum
efficiencies.

%
\end{abstract}

\maketitle
\section{Introduction}
 \label{intro}

In recent years, several innovative concepts have been proposed for developing third-generation
photovoltaic solar cells of high efficiency. As examples, multi-junction cells maintain the world record on
conversion efficiency, exceeding 40\% \cite{guter09}. However, their
commercial production is severely limited by the complexity of the device fabrication
process. Intermediate-band solar cells (IBSCs) \cite{Luque97} provide an intuitive
approach to significantly increasing photovoltaic conversion efficiency in a
single-junction solar cell device. A properly located intermediate band (IB) in the intrinsic
band gap serves as a ``stepping stone" in allowing photons with energies below the band gap to excite
electrons from the valence band (VB) to the conduction band (CB) via a two-step process. Through
these additional excitation channels, the lower-energy photons in the solar spectrum are able to contribute to the photocurrent
as well, resulting in a maximum efficiency of 63.2\%, substantially higher than the Shockley-Queisser
limit of 40.7\% for single band-gap solar cells \cite{Shockley61}. Since its conception, several key ingredients of the concept have been convincingly established \cite{Marti06, Luque05, Lopez11}, and much effort has been devoted to exploring different aspects of IBSCs \cite{Popescu08, Luque10, Bailey11, Yu08, Wang09, Lee10, Peng03, Keevers94, Waver94}. For example, a multi-intermediate band structure has been proposed to extend the system from containing one intermediate level to a number of intermediate bands, and the resulting efficiency limit can be more than 80\%\cite{Peng03}. To realize an IBSC, some specific material systems have been proposed, such as quantum dots\cite{Marti06, Luque05, Popescu08, Luque10, Bailey11}, semiconductor alloys\cite{Lopez11, Yu08, Wang09, Lee10}, semiconductor superlattices\cite{Peng03}, and dopant impurities\cite{Keevers94, Waver94}; yet to date, none of those materials could deliver the high efficiencies as expected from the high limits of IBSCs. In these efforts, one standing challenge is how to controllably build IBs in the intrinsic band gaps of candidate materials.

Recently, a new approach, termed non-compensated n-p codoping, has been proposed to create one or more tunable IBs in wide band-gap oxide semiconductors, as demonstrated for TiO$_2$ \cite{zhu09}, a material with a variety of desirable properties. In this approach, an n-type dopant contributes electrons to the host material, and a p-type dopant accepts electrons from the host; consequently, both the thermodynamic and kinetic solubilities of the dopants are enhanced by the Coulomb attraction between the oppositely charged n-p dopant pairs. In particular, by choosing the n and p dopants to possess different charge states, their non-compensated nature will ensure the creation of the IB. Here we note that the position and intensity of the IBs can also be tuned by choosing different combinations and concentrations of the non-compensated n-p dopant pairs. As for the material choice, TiO$_2$ has been considered as one of the most promising candidates for solar energy utilization, for example as photocatalysts in hydrogen production via water splitting \cite{Fujishima72, Asahi01, Khan02} and as charge collectors in dye-sensitized solar cells \cite{Regan91, Bach98}, because of its low cost, chemical inertness, photo-stability, and excellent charge transport properties \cite{Fujishima08}.

Since the non-compensated n-p codoping approach was proposed, it has been exploited both theoretically and experimentally in an increasing number of systems that demand precise dopant control for property optimization \cite{Pan10,Wang10,Liu11,Zhang11,Mangham11,Ma11,Hamal11}. For example, an \emph{ab initio} study of Cr-O codoped GaN showed an emergent IB and a controllably narrowed band gap with enhanced carrier mobility and photocatalytic activity in the visible light region \cite{Pan10}. Experimentally, Co-(C,S) codoped TiO$_2$ and Fe-N codoped TiO$_2$ exhibited enhanced dopant solubility and photocatalytic activity, thereby confirming the advantage of this approach \cite{Mangham11,Hamal11}.

In this paper, we shift our attention from the effects of the non-compensated approach on detailed electronic structures of solar cell materials, to the enhancements in the quantum efficiency of the solar cells based on these materials. We propose to develop IBSCs by exploiting the very existence of the IBs in such non-compensated n-p codoped materials, and explore theoretically their maximum quantum efficiencies under two different design schemes. We first focus on codoped TiO$_2$, and show that the maximum efficiency of the corresponding ideal IBSC can reach 52.7\%. However, in the original, idealized scheme, a small deviation of the IB position from the optimum value would cause a large drop in the quantum efficiency. To relieve this stringent yet undesirable requirement, we propose a new design scheme where current is also extracted from the IB, taking advantage of the delocalized nature of the IB built via non-compensated n-p codoping \cite{Mannella}. The IB position can now be located in a broader range within the intrinsic band gap for sufficiently high quantum efficiencies, with the maximum efficiency increased to 56.7\%. These findings suggest that the second design scheme should be more desirable in facilitating practical implementation of IBSCs.

The remainder of this paper is organized as follows. In \Sec{npcodoping}, we briefly review the central ingredients of the non-compensated n-p codoping approach as it is applied to Cr-N codoped TiO$_2$. In \Sec{model}, we present a general description of the IBSC model, including the four ideal conditions invoked, and obtain the necessary equations describing the electron transitions between the different energy bands. In \Sec{result}, we exploit the quantum efficiency of non-compensated n-p codoping TiO$_2$ as IBSC materials under two different design schemes, and also discuss the efficiency improvement in more general cases. In \Sec{summary}, we discuss the range of applicability of the models considered and summarize the main findings.

 \section{non-compensated n-p codoping}
 \label{npcodoping}
Non-compensated n-p codoping has been proposed as an appealing concept for controlled tuning of the band gap of TiO$_2$ and other wide band-gap semiconductors with potentially much improved photoactivity and functionality \cite{zhu09}. Earlier attempts use compensated dopants \cite{Gai09}, i.e., the number of electrons introduced by an n-type dopant equals the number of holes introduced by a p-type dopant; therefore, the dopants largely compensate each other as they form a local pair in the host semiconductor. In contrast, in the non-compensated n-p codoping approach, an n-p dopant pair is ensured to contribute net charge carriers (electrons or holes) to the host. This new approach embodies two key elements: (1) the Coulomb attraction within the n-p codopant pair enhances both the thermodynamic and kinetic solubilities of the dopants; (2) the non-compensated nature ensures the creation of impurity/intermediate bands in the gap region of the host semiconductor. Furthermore, the position and intensity of the IBs can be tuned by different combinations and concentrations of the non-compensated n-p pairs; the hybridization between the n and p dopants will broaden the impurity levels into more extended IBs.

The anatase phase of TiO$_2$ is known to be more reactive in water splitting \cite{Fujishima08,Chen10}, and when codoped with the non-compensated Cr-N pairs, an IB is created in the intrinsic band gap, which is effectively narrowed. This narrowing, in turn, enables TiO$_2$ to absorb the more abundant visible light, rather than the ultraviolet region of the sunlight alone. The Cr-N pair has also been shown to encounter lower kinetic barriers in reaching substitutional sites. Moreover, strong hybridization between the Cr $3d$ and N $2p$ orbitals produces the broadened IB in contrast to the highly localized impurity levels contributed by Cr or N doping alone, a feature more desirable for efficient electron-hole separation, because the photo-generated carriers have high mobility in the IB as revealed in recent experiments \cite{Mannella}. These energetic features, especially the delocalized IB in the Cr-N codoped TiO$_2$, provide the basis for the two different design schemes of the IBSCs.

 \section{Model description and ideal conditions of IBSC}
 \label{model}

\begin{figure}
\includegraphics[width= 3in]{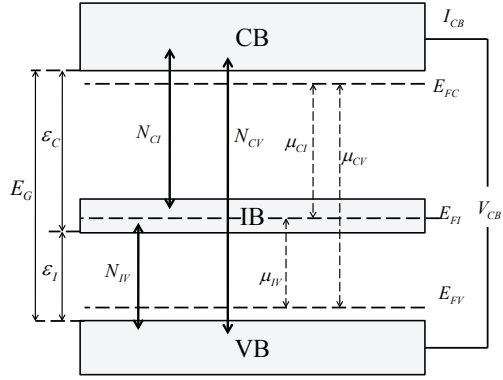}
\caption{\label{fig1} Band diagram of an intermediate-band solar cell. $E_G$ is the intrinsic energy band gap, $\varepsilon_I$ is the IB position, and $\varepsilon_C=E_G-\varepsilon_I$ is the gap between the IB and CB. $E_{FV}$, $E_{FI}$ and $E_{FC}$ are the quasi-Fermi levels for the three bands. Chemical potentials $\mu_{IV}$, $\mu_{CI}$ and $\mu_{CV}$ are the spacings between the three quasi-Fermi levels. $N_{IV}$, $N_{CI}$, and $N_{CV}$ describe the net numbers of electron transitions between the different energy bands. $I_{CB}$ is the current extracted from the CB at the voltage of $V_{CB}$.}
\end{figure}

We consider here an IBSC model containing the VB, the CB, and the IB located between them. The band diagram and electron transitions between the bands are schematically shown in \Fig{fig1}. As adopted in previous works on the quantum efficiency of IBSCs \cite{Luque97,Peng03}, our approach is also based on the Shockley and Queisser (SQ) model \cite{Shockley61}.
For convenience, here we reproduce four of the seven ideal conditions in the original IBSC model \cite{Luque97} that are especially pertinent to the present work: (C1) full absorption of photons whose energies are sufficient to induce electronic excitations between any two of the three bands; (C2) nonradiative electronic transitions are forbidden; (C3) no current is extracted from the IB unless otherwise differently specified; (C4) for every range of energies only one of the three absorptions is important (namely, the ranges of energies that induce the three electronic excitations do not overlap with each other).

Between different energy bands, electron-hole (e-h) pairs can be generated or annihilated associated with the absorption or emission of photons. The balance of the net electrons transferred gives the current output. Note that in this work the number of electron transitions per unit time and per unit illuminated area is denoted as $N$, and the current $I$ is given by the relation of $I=e\cdot N$, where $e$ is the electron charge. According to C1, C2, and C4, the rate of the e-h pair generation (annihilation) per unit illuminated area is equal to the flux of absorbed (emitted) photons in the corresponding energy interval. The latter is related to the Bose-Einstein distribution function as\cite{Landau67}:
\be \label{b-e}
\mathcal{N}(\varepsilon_m,\varepsilon_M, T, \mu)=\frac{2\pi}{h^3c^2}\int_{\varepsilon_m}^{\varepsilon_M}\!\!
\frac{\varepsilon^2d\varepsilon}{e^{(\varepsilon-\mu)/k_BT}-1},
\ee
where $T$ is the temperature, the energy interval is $\varepsilon_m<\varepsilon<\varepsilon_M$, $\mu$ is the local chemical potential or the quasi-Fermi-level separation of the energy bands in thermodynamic equilibrium\cite{Peng03}, $k_B$, $c$ and $h$ are the Boltzmann constant, the speed of light, and the Planck constant, respectively.

The quantity that comes into play is the rate of the e-h pair generation minus that of the e-h pair annihilation; this quantity gives the net number of electron transitions per unit time. As shown in \Fig{fig1}, the net numbers of electron transitions from VB to IB, IB to CB, and VB to CB per unit time and per unit illuminated area are given respectively by:
\begin{eqnarray}
&\scalebox{0.9}{$N_{IV}=\begin{cases}\mathcal{N}(\varepsilon_I,\varepsilon_C,T_s,0)-\mathcal{N}(\varepsilon_I,\varepsilon_C,T_a,\mu_{IV})& \text{ $\varepsilon_I<E_G/2$}\\\mathcal{N}(\varepsilon_I,E_G,T_s,0)-\mathcal{N}(\varepsilon_I,E_G,T_a,\mu_{IV})& \text{ $\varepsilon_I\geq E_G/2$}\end{cases}$},&\nonumber
\\&\scalebox{0.9}{$N_{CI}=\begin{cases}\mathcal{N}(\varepsilon_C,E_G,T_s,0)-\mathcal{N}(\varepsilon_C,E_G,T_a,\mu_{CI})& \text{ $\varepsilon_I<E_G/2$}\\\mathcal{N}(\varepsilon_C,\varepsilon_I,T_s,0)-\mathcal{N}(\varepsilon_C,\varepsilon_I,T_a,\mu_{CI})& \text{ $\varepsilon_I\geq E_G/2$}\end{cases}$},&\nonumber
\\&\scalebox{0.9}{$N_{CV}=\mathcal{N}(E_G,\infty,T_s,0)-\mathcal{N}(E_G,\infty,T_a,\mu_{CV})$}.\qquad \qquad \qquad  &
\label{net-e}
\end{eqnarray}
Here $E_G$ is the intrinsic energy band gap, $\varepsilon_I$ is the IB position, $\varepsilon_C=E_G-\varepsilon_I$ is the gap between the IB and CB, $\mu_{IV}$, $\mu_{CI}$ and $\mu_{CV}$ are the chemical potentials or spacings between the three quasi-Fermi levels, $T_s$ and $T_a$ is the temperature of the sun and IBSC, respectively. Note that the chemical potential of photons has been taken to be zero. On the right hand side of each line of \Eq{net-e}, the first term represents the absorption of photons that generates e-h pairs, and the second term the radiative emission of photons that annihilates e-h pairs. Here we take the transitions between the VB and CB as a specific example. Photons absorbed by the VB $\rightarrow$ CB transitions or emitted by the CB $\rightarrow$ VB transitions have energies equal to or larger than $E_G$. Thus, the rate of e-h pair generation is given by $\mathcal{N}(E_G,\infty,T_s,0)$, while the rate of e-h annihilation is $\mathcal{N}(E_G,\infty,T_a,\mu_{CV})$. Therefore, the net number of electron transitions from the VB to CB is given by the difference $\mathcal{N}(E_G,\infty,T_s,0)-\mathcal{N}(E_G,\infty,T_a,\mu_{CV})$, defined by $N_{CV}$ in \Eq{net-e}. Similarly, $N_{IV}$ ($N_{CI}$) is obtained as the net number of electron transitions from the VB to IB (from the IB to CB), with a proper handling of the energy interval according to assumption C4.

The current delivered to the external load is determined by the balance of the net electrons transferred, i.e., $N_{IV}$, $N_{CI}$, and $N_{CV}$. For example, if we look at the CB, the current returning to the CB is given by $I_{CB}/e=N_{CI}+N_{CV}$. In the next section, we will discuss the output current and the quantum efficiency of the IBSCs under the two different design schemes.

 \section{Quantum efficiency of IBSC under two design schemes}
 \label{result}

\subsection{Quantum efficiency of TiO$_2$ as IBSC
    without current extraction from the IB}
\label{resultA}

\begin{figure}
\includegraphics[width= 3in]{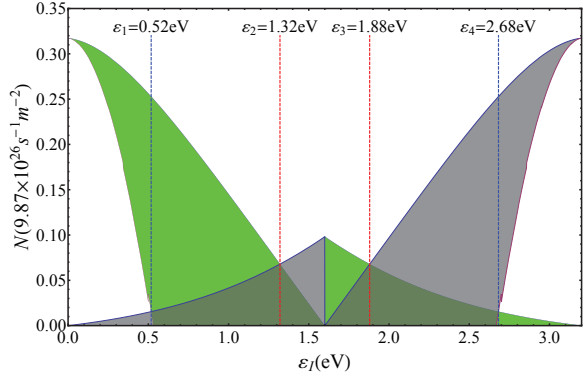}
\caption{\label{fig2} (Color online) Range of $N_{IV}$ (green area) and $N_{CI}$ (gray area) defined in \Eq{net-e}.}
\end{figure}

First, we consider that there is only current extracted from the CB and the number of electrons in the IB is conserved. When the IB is quite close to the VB, photons that can induce the VB $\rightarrow$ IB transitions are much more than those that can induce the IB $\rightarrow$ CB transitions. As a result, $N_{IV}$ defined by \Eq{net-e} is calculated to be always larger than $N_{CI}$ if $\varepsilon_I < \varepsilon_1=$ 0.52 eV, as shown in \Fig{fig2}. Here $\varepsilon_i (i=1,2,3,4)$ are the energy values where the extremum of $N_{IV}$ equals that of $N_{CI}$. In contrast, $N_{IV}$ is always smaller than $N_{CI}$ if the IB is quite close to the CB ($\varepsilon_I > \varepsilon_4=$ 2.68 eV). The lack of overlap between the ranges of $N_{IV}$ and $N_{CI}$ for  $\varepsilon_I < \varepsilon_1$ and  $\varepsilon_I > \varepsilon_4$ is in contradiction with the balance of electrons in the IB, indicating that the ideal conditions cannot be satisfied in these cases. For $\varepsilon_I < \varepsilon_1$, the net electron transitions between the VB and IB are suppressed, as they must match the transitions between the IB and CB. The suppression can be caused by several deviations from the ideal conditions, such as partial absorption of the photons that induce the VB $\rightarrow$ IB transitions (violation of C1), or nonradiative transitions from the IB to VB (violation of C2). Similarly, for $\varepsilon_I >  \varepsilon_4$, electron transitions from the IB to CB are suppressed and the condition C1 cannot be satisfied. For the above two cases, we assume that the smaller one of $N_{IV}$ and $N_{CI}$ dominates the two-step transitions from the VB to CB and some non-ideal factors are considered in the other step. For $\varepsilon_1 \leq \varepsilon_I \leq \varepsilon_4$, there are overlaps between the ranges of $N_{IV}$ and $N_{CI}$, and the ideal conditions can be satisfied. As a consequence, the current from the CB is obtained as:
\be \label{ICB}
I_{CB}/e=\Bigg\{\begin{array}{cc}
     N_{CI}+N_{CV} \qquad \qquad \qquad \quad \!\!\!\! & \varepsilon_I < \varepsilon_1 \\  N_{CI}+N_{CV}, \text{with } N_{IV}=N_{CI}  & \varepsilon_1 \leq \varepsilon_I \leq \varepsilon_4
     \\  N_{IV}+N_{CV} \qquad \qquad \qquad \quad \!\!\!\! & \varepsilon_I > \varepsilon_4
  \end{array}.
\ee
This current is delivered at a voltage $V_{CB}=\mu_{CV}/e$. The quantum efficiency of the IBSCs is given by:
\be \label{eta1}
\eta=\frac{I_{CB}V_{CB}}{\sigma T_s^4},
\ee
where $\sigma$ is the Stefan-Boltzmann constant.

The optimum quantum efficiency as a function of the IB position is presented by the black line in \Fig{fig3}. Note that in all of our numerical calculations, $E_G$ is set to 3.2 eV for the band gap of TiO$_2$, $T_s$= 6000 K, and $T_a$  = 300 K. The maximum quantum
efficiency, 52.7\%, can be obtained given $V_{CB}=$ 3.00 V and  $\varepsilon_I=$ 1.32 eV or 1.88 eV where $N_{IV}$ and $N_{CI}$ are well matched as shown in \Fig{fig2}. The efficiency curve has two symmetrical peaks centered on the mid-gap $\varepsilon_I= \varepsilon_C=E_G/2=$ 1.6 eV, a reflection of the e-h symmetry in the IB around this location. Since the two peaks are sharp, a small deviation in the IB position from the optimum value would lead to a large drop in the quantum efficiency. This large drop could be partly responsible for the observation that the potential high efficiency of IBSCs has not been achieved experimentally, i.e., the IB would have to be built very close to the optimum position.

\subsection{Quantum efficiency of TiO$_2$ as IBSC
    with current extraction from the IB}
\label{resultB}

To relax this stringent working condition for IBs, we explore a new design scheme by extracting current also from the IB, taking advantage of the delocalized nature of the IB in the non-compensated n-p codoped system. Another motivation for this consideration is the observation that $N_{IV}$ can be much larger than $N_{CI}$ when $\varepsilon_I <\varepsilon_2=$ 1.32 eV (\Fig{fig2}); however, if there is no current extraction from the IB, the large value of $N_{IV}$ cannot be fully utilized. Extracting current from the IB can make a better use of the possible high values of $N_{IV}$, thus improving the efficiency. In this scheme, the IB should be connected to a charge collecting contact that is isolated from the contact of the CB. Here we note that a recent experimental study of IBSC has demonstrated the importance and enhanced efficiency of the separation (or blocking) between the IB and CB, even though no attempt was made to extract current directly from the IB \cite{Lopez11}.

\begin{figure}
\includegraphics[width= 3in]{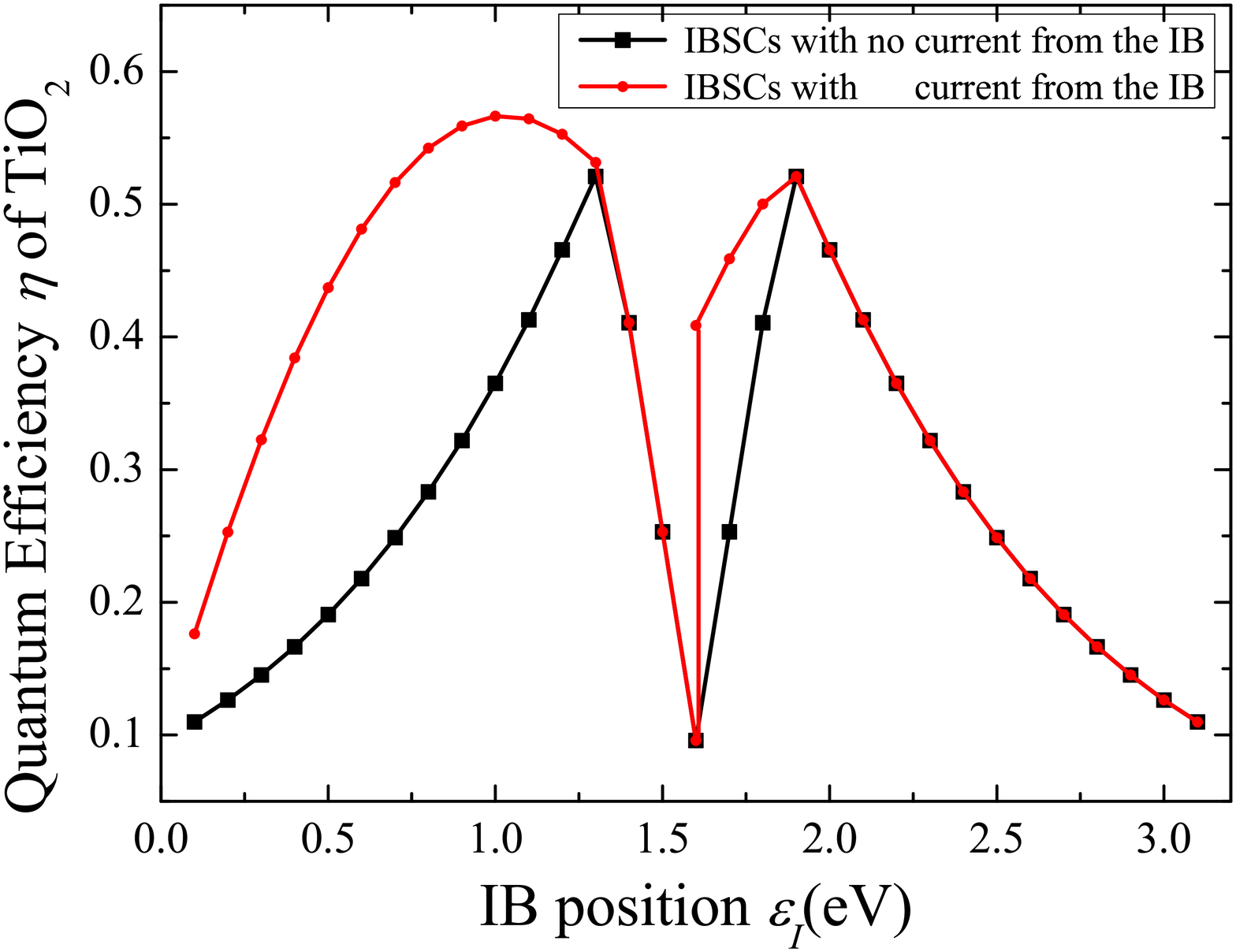}
\caption{\label{fig3} (Color online) The optimum quantum efficiency of TiO$_2$ as IBSCs without (black line) and with (red line) current extraction from the IB as a function of the IB position.}
\end{figure}

We assume that the net number of electron transitions between two of the three bands are still determined by \Eq{net-e}, and consider the case that only electrons can be extracted from the IB. The currents from the IB and CB are given respectively by:
\begin{subequations}
\begin{eqnarray} \label{IIBCB}
 I_{IB}/e &=& N_{IV}-N_{CI} \geq 0, \\
 I_{CB}/e &=& N_{CI}+N_{CV}.
 \label{IIBCB2}
\end{eqnarray}
\end{subequations}
$I_{IB}$ and $I_{CB}$ are delivered at two different voltages, $V_{IB} = \mu_{IV}/e$ and $V_{CB} = \mu_{CV}/e$. The quantum efficiency in this new scheme is revised from \Eq{eta1} to:
\be \label{eta2}
\eta=\frac{I_{IB}V_{IB}+I_{CB}V_{CB}}{\sigma T_s^4}.
\ee
In \Fig{fig3}, the red line represents the optimum quantum efficiency of TiO$_2$-based IBSCs as a function of the IB position,
with current extracted from the IB. The maximum quantum efficiency can reach 56.7\% with  $\varepsilon_I=$ 1.03 eV, $V_{CB}=$ 3.00 V and $V_{IB}=$ 0.97 V. It also exhibits the obvious double-peaked feature,
but the two peaks are no longer symmetric, since the IB can only
output electrons. Most notably, in the vicinity of  $\varepsilon_I=$ 1.03 eV, the efficiency within the present design scheme can be substantially higher than that of the ideal one, and the stringent requirement of the IB position is relaxed.
When the IB is quite close to the CB, e.g., $\varepsilon_I>\varepsilon_4=$ 1.88 eV, photons that can induce the IB
$\rightarrow$ CB transitions are abundant, and all the electrons excited to the IB from the VB can be readily excited to the CB; therefore, there is no need to extract current from the IB in this regime (see \Fig{fig3}). The jump in the quantum efficiency at the mid-gap ($\varepsilon_I=\varepsilon_C=E_G/2=$ 1.6 eV) is due to the assumed condition C4. If the overlap between the three absorption coefficients \cite{Cuadra04, Navruz08} is considered, the jump, or discontinuity, can be removed, with the main results staying valid.

\subsection{Quantum efficiency of general IBSC
   without and with current extraction from the IB}
\label{resultC}

So far, our quantitative analysis has been focused on the specific system of Cr-N codoped TiO$_2$. Now we broaden our attention to other related systems, such as the Cr-O codoped GaN system \cite{Pan10}. Without loss of generality, we look into the efficiency improvement of IBSCs (the codoped TiO$_2$ and GaN are the subset systems of this generic consideration). For scheme I with no current extraction from the IB, the current delivered at a voltage $V_{CB}$ can be given by \Eq{ICB} together with \Eq{net-e}. The corresponding quantum efficiency of the IBSCs is again obtained from \Eq{eta1}. For scheme II with current extraction from the IB, the currents from the IB and the CB are given by \Eqs{IIBCB} and (\ref{IIBCB2}), and the voltages are $V_{IB}$ and $V_{CB}$ respectively. The total quantum efficiency can be obtained from \Eq{eta2}.

\begin{figure}
\includegraphics[width= 3in]{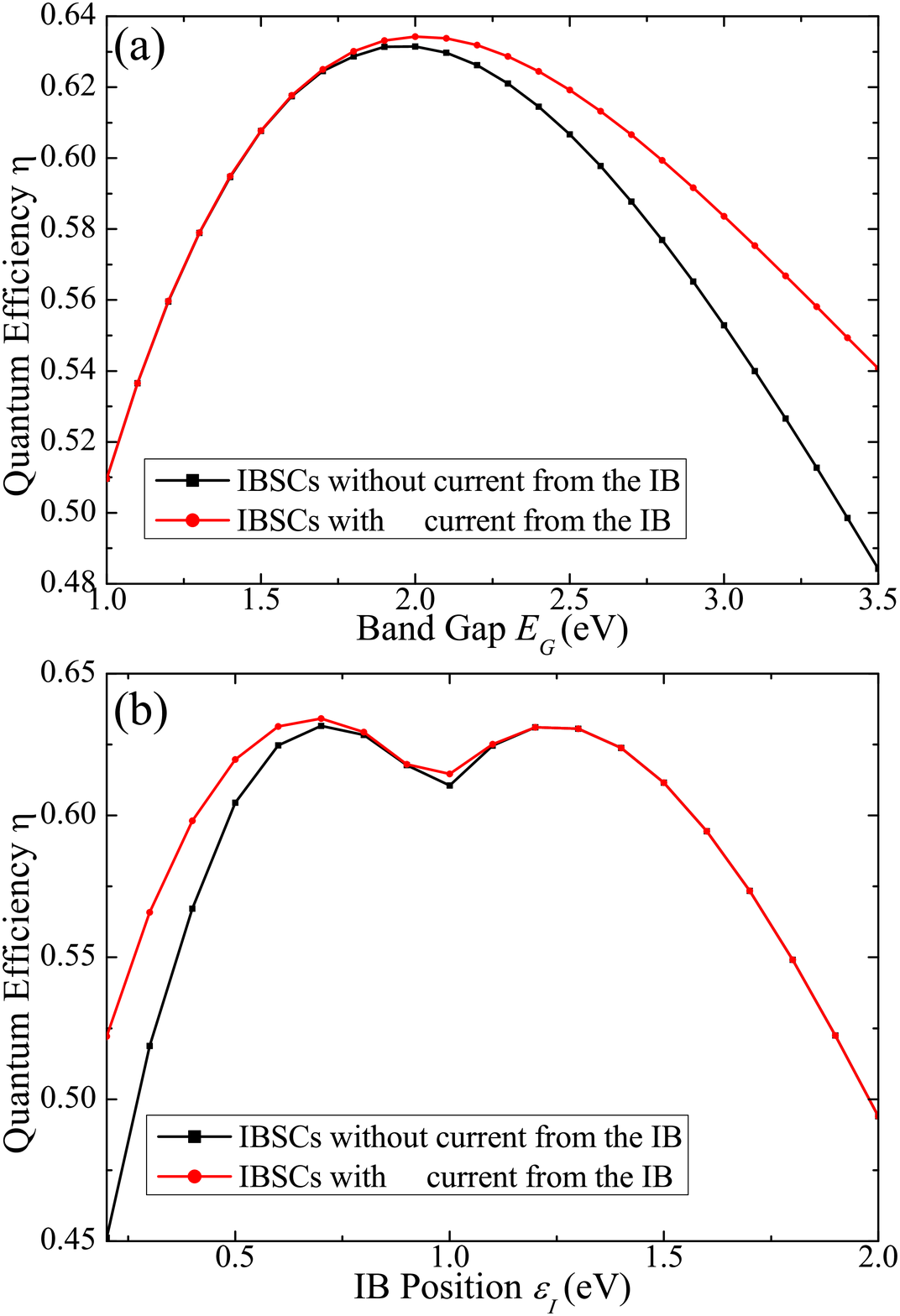}
\caption{\label{fig4} (Color online) The optimum quantum efficiency of general IBSCs without (black line) and with (red line) current extraction from the IB versus (a) the band gap and (b) the IB position.}
\end{figure}

Figure 4 shows the optimum quantum efficiency of general IBSCs without and with current extraction from the IB versus the band gap or the IB position. To obtain these curves, we first select an $\varepsilon_I$ and continuously change the value of $E_G$ from $\varepsilon_I$ to an upper bound (in our calculation this upper bound is set to 4.0 eV since almost all the band gaps of typical semiconductors are less than 4.0 eV), and then calculate the efficiencies. The maximum efficiency as a function of $E_G$ is shown in \Fig{fig4}(a), and the maximum efficiency as a function of $\varepsilon_I$ is shown in \Fig{fig4}(b). The maximum efficiency of the IBSCs under the new design scheme can reach 63.4\% with $E_G=$ 2.02 eV
and $\varepsilon_I =$ 0.69 eV, which is only slightly higher than the maximum efficiency of 63.2\% without current extraction from the IB. However, as shown in \Fig{fig4}(a), extracting current from the IB can significantly improve the efficiency for wide band-gap materials ($E_G >$ 2 eV), suggesting that the present new design scheme is more promising for oxide-based IBSCs. Figure 4(b) also shows that, in agreement with the results for TiO$_2$, there is also no need to extract current from the IB of general IBSCs when the IB is close to the CB.

\section{Discussion and Summary}
\label{summary}

In this section, we will briefly discuss the merits and disadvantages associated with the different design schemes of IBSCs based on the non-compensated n-p codoping materials.

The non-compensated n-p codoping approach is highly promising in tuning the band structures of oxide materials and related wide band-gap semiconductors, as reflected by the dramatically enhanced absorption of photons in the visible light region for TiO$_2$. On the other hand, the codoping scheme naturally introduces more impurity atoms into the host semiconductors, which might serve as recombination centers for the e-h pairs, potentially shortening the exciton lifetime and mobility. Furthermore, on a practical level, nonradiative recombination (such as scattering with phonons) can also influence the efficiency of photovoltaic conversion in non-compensated n-p codoped TiO$_2$. Nevertheless, the present theoretical calculations under the idealized conditions allow us to explore the potential maximum quantum efficiency of the IBSCs and serve as guidance in future design of IBSCs.

Another important issue in IBSCs is how to achieve current-matching. We have examined two different design schemes: 1) a general setup with two terminal electrodes for charge collecting and pumping; 2) a new design with a third electrode for extracting current from the IB. In the new design, two isolated contacts are needed to collect charge from the CB and IB in parallel. The realization of this design is similar to the implementation of energy selective contacts (ESCs) for hot carrier solar cells\cite{Ross82,Shrestha10}. These ESCs, like an energy filter, only allow carriers within a narrow energy range to pass through to the contacts, and carriers with other energies are reflected back. Such ESCs matching the energy range of IB can be used as the isolated contact for the third electrode in the IBSCs in our new design. Of course, there are more practical issues to be considered in real devices. Moreover, detailed device realization of the different IBSC designs using the codoped TiO$_2$ materials will surely require more sophisticated engineering approaches, and may need to compare the advantages and limitations using material structures in thin-film or nanowire/nanotube form. In this present work, we mainly introduce the novel materials into the field of IBSC with the objective of broadening the material choices beyond what have been considered so far.

In summary, we have explored the optimum quantum efficiency of IBSCs based on codoped TiO$_2$ under two
different design schemes. When the ideal conditions are preserved, the corresponding maximum quantum
efficiency for the codoped TiO$_2$ can reach 52.7\%, but requires a stringent IB position. Upon extracting current
from the IB, the IB position can be in a wide range,
with a maximum efficiency of 56.7\%. These results should facilitate experimental realization of IBSCs, and make
n-p codoped TiO$_2$ as appealing candidate materials for high-efficiency solar energy utilization. In potential future realization of those design schemes using non-compensated n-p codoped wide band-gap semiconductors, various pressing technical or practical challenges will have to be resolved.

\begin{acknowledgments}
 We thank Tianli Feng, Guangwei Deng, and Yi Xia for valuable discussions. This work was supported by the National Natural Science Foundation of China (Grant No. 11034006), the Fundamental Research Funds for the Central Universities (Grant Nos.2340000029 and 2340000024), and in part by Division of Materials Sciences and Engineering, Basic Energy Sciences, US Department of Energy (Grant No. DE-FG03-02ER45958).
\end{acknowledgments}

\nocite{*}

\end{document}